# Rigorous Calculation of the Partition Function for the Finite Number of Ising Spins

Alexey A. Peretyatko, Ivan A. Bogatyrev, Vitaliy Yu. Kapitan[a)], Yury V. Kirienko,
Konstantin V. Nefedev[b)], Valery I. Belokon
Far Eastern Federal University
School of Natural Science
Russia, Vladivostok, Sukhanova str.8-43
a) kvy@live.ru
b) knefedev@phys.dvgu.ru

**ABSTRACT**

The high-performance scalable parallel algorithm for rigorous calculation of partition function of lattice systems with finite number Ising spins was developed. The parallel calculations run by C++ code with using of Message Passing Interface and massive parallel instructions. The algorithm can be used for the research of the interacting spin systems in the Ising models of 2D and 3D. The processing power and scalability is analyzed for different parallel and distributed systems. Different methods of the speed up measuring allow obtain the super-linear speeding up for the small number of processes. Program code could be useful also for research by exact method of different Ising spin systems, e.g. system with competition interactions.

**Keywords:** Massive-Parallel Algorithm, MPI, C++, Parallel and Disributed Calculations, Ising Model, Partition Function, Statistical Mechanics.

## 1. INTRODUCTION

Among the huge variety of fundamental scientific tasks there are problems from the solution of which ones the comprehension progress of the Nature depends in whole. The solution of Ising model, a simple mathematical model of phase transitions of correlated systems, could be such a problem. The rigorous solution in this model it is necessary not only for the development of a statistical physics, but also and in related areas: nanotechnology, nanomaterial science, nanobiotechnology and et al. Today it is possible to work within systems which ones consist of relative small (finite) number of atoms. The exact solution in Ising model would be very desirable for further development of science and as consequence of a technology. The existing today single-threaded searching methods have strong limitations and not allow obtain the solution within a reasonable period of time. The total amount of a terms partition function grow up as power function with the increasing of the number of elements in the system, moreover Ising model could not be solved exactly by analytical approach in 3D, since the non-planar tasks are NP-complete problem [1-2]. Thus for exact construction of the partition function even for small finite systems it is desirable have a scalable algorithm for the using of computer power which one could support the intensive parallel calculations. The productivity of computers enhances every ten years approximately in thousand times, that essentially expands the class of solvable numerical tasks.

Approximate research methods of 2D and 3D Ising model, such for example as Metropolis algorithm and method Monte Carlo (MC) [3-4] have wide acknowledgement. With respect to rigorous solution the situation is more problematical. The rigorous analytical solution for 1D case was obtained by Ising in 1925, and exact solution for 2D case was obtained by Onsager in 1944 [5] (difference between "exact" and "unrigorous" defined in [6]). In work [7] by authors it was shown that the critical temperatures of transition of finite systems to ferromagnetism could be calculated more precisely than in approximate analytical theories. Therefore it is interesting to construct the rigorous solution for finite number of spins for a large lattices with the number of nearest neighbors Z=4 or more, and make the comparison of results, which ones can be obtained by means of numerical simulation and method rigorous calculation.

The aim of this work is development, optimization and testing on a different computational resource of high-performance scalable parallel algorithm of rigorous calculation of partition function for the finite number $N$ of Ising spins with ferromagnetic exchange on the planar lattice with Z=4, and show the possibility of generalization of this algorithm on 3D arbitrary lattices.

This work was done at support of the Russian Federal Aim Program "Scientific and scientific-pedagogical staff of innovation Russia", Ministry of Education and Science of Russian Federation, State Contract № 14.740.11.0289, "Development of computer programs based on algorithms of parallel programming and optimization of high-performance distributed computing environment for solving problems of natural science."

## 2. ISING MODEL

Let us shortly remind the main idea of the simplest model of ferromagnetism - the Ising model, which one has not rigorous solution yet in spite of all its simplicity. In this model it permits only the interactions between nearest neighbors usually. The Hamiltonian of the system in general case is

$$H = -\frac{1}{2}\sum_{i=1}^{N}\sum_{j=1}^{Z} J_{ij} S_i S_j - h \sum_{i}^{N} S_i, \qquad (1)$$

where $h$ is external magnetic field, the summation over $j$ denotes the summation over neighbors (nearest here). Using summands in the Eq. (1) it is possible to put for the each configuration of $2^N$ possible configurations in correspondence two numbers – the energy $E_i$ and the spin excess $M_i$ [8]. Exchange constant $J_{ij}$ is positive and in considered case equals to one (in principle could be sign-changing and distance-

sensitive in general case). The partition function of finite number of Ising spins

$$Z_N(h,T) = \sum_{S_1}\sum_{S_2}\ldots\sum_{S_N} Exp\left[-\frac{H}{T}\right]. \quad (2)$$

The troubles with the rigorous numerical calculation of Eq. (2) are in the estimation of the degeneracy multiplicity of the spin excess $M_i$ (total sum of all $S_i$) over energy $E_i$ (total sum exchange energy of all pair spins). The degeneracy means, that to one value of $M_i$ it could correspond a different values of $E_i$. The degeneracy multiplicity equals to the number of combinations with given quantity of pairs of side by side "units" ("zeros") with taking into account the periodical boundary conditions. In other words in the combinatorial terms we need estimate the number of permutations with repeating and limitations on the distribution of side by side elements. The total amount of the degeneracy multiplicities for the fixed $M_i$ over all permitted $E_i$ is the binomial coefficient.

## 3. SCALABLE PARALLEL ALGORITHM

The simplicity of proposed approach gives the possibilities of extension of the elaborated algorithm for 2D Ising lattice, also and on 3D lattice with given number of a nearest neighbors $Z$. The limitation of linear search parallel algorithm is in linear dependence from computer performance.

In general case the lattice can be presented by the set of bit-vectors. For 2D this set bit-vectors $\{a_i\}$ could be the one dimension array, where two nearest elements correspond in two nearest columns of lattice, for 3D it is two dimensional matrix $\{a_{ij}\}$. De facto the information about distribution of spin states each column of the lattice was written it the short unsigned integer variables, number of which ones equals to linear dimension of lattice. Integer variable size is four bytes, i.e. 32 bits. For calculations of a small lattice it needs small number of these bits, hereinafter referred to as significant. A unary and pair logical as well as arithmetical operations with a parts of such bit-vectors, the number of significant bits which ones equal to second linear dimension of lattice, allow obtain the $M_k$ and $E_k$. Total sum of $M_k$ and $E_k$ of all variables allow obtains the $M_i$ and $E_i$ for one configuration. All configurations can be "counted" by parallel changing of value of each integer variable in a diapason from 0 to 2 in power $n$, where $n$ is a linear dimension of the lattice.

### a. Spin Excess

The sum of all units of all bit-vectors gives the number of spins, which ones have direction "up". It is easy to define the spin excess as the deference between the total numbers of spins and the double amount of the spin "up". The quantity of units was defined with using of Kernighan algorithm [9]

```
int q_units(unsigned short int i)
{ int j = 0;
  while (i)
  {i&=i-1; ++j;}
  return j; }
```

Really, in code of program it has used the optimization which one connects with the storage in an unsigned short integer cache array $\{b_i\}$ of the quantities units (weights of bit-word) calculated once for the each decimal numbers and reference to the elements of the array in cycles. The reference on the elements of the given array is used instead of an all arithmetical operations, where the enumeration of the bit-vector units takes place.

### b. Rows Exchange Energy

The calculation of exchange energy between spins in rows can be done in three steps:

*1)* The operation XOR between all pairs $a_i$ and $a_{i+1}$, including periodical boundary conditions $a_1 \wedge a_N$, for 2D lattice ($a_{ij} \wedge a_{i+1,j}$ or $a_{ij} \wedge a_{i,j+1}$ for 3D case);
*2)* The invert conversion of the previous operation result, i.e. implimentation of NOT logical operation;
*3)* The reset of bits which ones have not information about spin state of the system. The subtraction from resultant bit-vector $2^{32}-2^n$ (for expample order 32 is for "int" type of the bit-word size);
The quantity of units in the resultant bit-word gives us the number of the pair exchange interactions with positive sign in rows, and should be stored in the separate variable $En$.

### c. Columns Exchange Energy

The enumeration of the positive pair exchange spins interactions in the each $a_i$ ($a_{ij}$) column achieve by the following way:

*1)* To each decimal number "j" - index of the integer aray $\{c_j\}$ - put in correspondence the dicimal number $c_j$, the binary notation of which one is circular shift of significant bits of number "j";
*2)* The implemention of the operation XOR between all pairs of integer number $a_i$ and $c_j$, where $j= a_i$;
*3)* The invert conversion (NOT operation) of the previous operation result;
*4)* The reset of bits which ones have not information about spin state of the system;
*5)* The references on elements of the array $\{b_i\}$ is used for the determination of units quantity and summation with $En$;

The final value of $En$ is the total amount of the all pair exchange interactions with positive sign for one configuration. There are not troubles to calculate the total exchange energy of the spin pairs interactions by using of the simple arithmetic expression $E_{tot}=2(-En+N)$.

### d. Parallelization. MPI

The Massage Passing Interface (MPI) standard extends the possibility of computers and makes available parallel algorithmization that allows the solving of the tasks, the solution of which ones couldn't be obtained in the reasonable time early. The possibility of the parallel execution in our C++ code is implemented via the MPI libraries.

The number the nested loops equals to the length of the vector $\{a_i\}$, i.e. number bit-vectors. The task of the parallelization of the nested iterations requires the application of the technology of dynamical processes generation. In the current version of the program we did not used the dynamical parallelization, but there are some possibilities of the organization of a dynamical generation of the parallel processes. For example, the group communicators can be used.

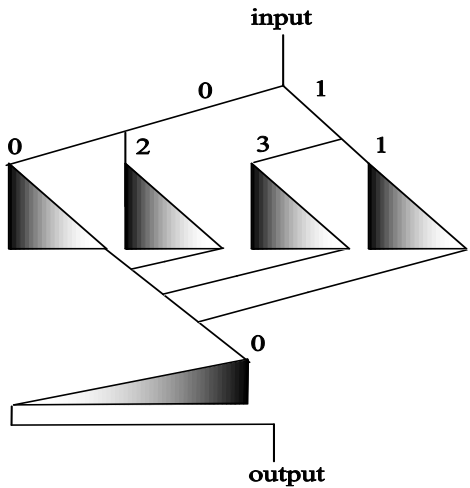

Figure 1. The logical scheme (graph) of the parallel algorithm with dynamical generation of two processes by two.

The communication process topology is on the Fig.1. The given approach is acceptable for the two nested loops

for(j=0; j<2;j++)
for(i=0; i<2;i++)

where the two processes were generated by two other. The data is distributed uniformly between parent processes, "0" and "1", and the child processes, "2" and "3". The "0" process is chosen as the root, and the calculation results are passed in it after completing of the computing. The root process receives, sorts and saves the resultant data.

## 4. TESTING THE CODE PERFORMANCE FOR DIFFERENT COMPUTER SYSTEMS

To find the fastest solution of the resource-intensive tasks supercomputers with multi-core architecture are used. However, such the machines are quite expensive and quickly become outdated without the possibility of upgrading. Usually the architecture is push-typed and has weak capability to modernization. It is possible to create cloud computing power with using of the regular workstations with heterogeneous architectures. Therefore, a heterogeneous, scalable computing system for the testing of the performance our algorithm was used. Also the work of program was checked on HP Cluster (16 and 32 cores).

### *GNU / Linux PelicanHP*

We used two computers: Console Cluster - Intel Core 2 Quad 2,5 Ghz, 4Gb DDR2, 500Gb HDD, 9800GTX 512Mb, 10/100/100 Mb Ethernet; cluster nodes: AMD Athlon 64 X2 2,8 GHz, 3 Gb DDR2, 232 Gb HDD, ATI Radeon HD 4850, 10/100/100 Mb Ethernet. The used GRID is the distribution GNU / Linux PelicanHPC x86_64 (v2.3.2) linux kernel 2.6.32-5, [10]; implementation MPI: OpenMPI v1.4.2-4. The loading of the operating system made from USB drive on the console of control computer. Service DHCP runs for a dynamic IP-addressing of the cluster nodes.

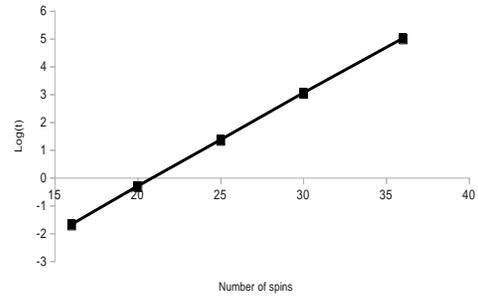

Figure 2. The dependence on logarithm of time (sec) from number of the spins.

The service runs NFS (Network File System) to create a shared resource across the network and to boot image is installed on Linux NFS, that one makes possible to run the cluster nodes over the network (PXE boot). After of the searching for available nodes the cluster is installed.

We run our program on 2 cores which ones is the ownership of two different physical processors at different motherboards connected directly to RJ-45 cable. On the Fig. 2 are the results of estimates of the computation time in dependence on the number of spins. According to the calculations it is possible to make prediction about the time of calculation of the 7x7 system. On 2 cores the time for system of 49 spins is about 76 years! It is easy to see that the scaling of the algorithm on more number of cores can give the solution in reasonable time.

### *HP Cluster*

To check the code performance it is used of HP cluster Intel Xeon E5410 @ 2.33GHz. The 16 cores have used for estimation of the executing time dependence from number of cores. There are time logarithms of program running (in sec.) on 16 cores for different numbers of spins (36 and 30 spins) on the Fig. 3. Unfortunately the available resources couldn't give us the possibilities to estimate of the solution time for bigger number of spins.

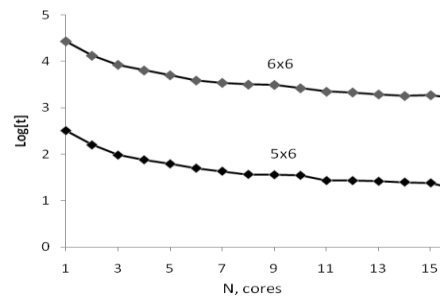

Figure 3. The dependence on logarithm of time from number of the cores for defferent number of spins.

The elaborated algorithm for the rigorous calculation of partition function for finite number of Ising spins on the square lattice displays the scalability. Processes are independent and run approximately the same time.

TABLE 1. Results of calculation for 4x4 system.

| Nc | $M_i$ | $E_i$ | Nc | $M_i$ | $E_i$ | Nc | $M_i$ | $E_i$ |
|---|---|---|---|---|---|---|---|---|
|   |   |   | 64 | 2 | -12 |   |   |   |
| 1 | 16 | -32 | 624 | 2 | -8 | 96 | -4 | -12 |
| 16 | 14 | -24 | 1920 | 2 | -4 | 704 | -4 | -8 |
| 32 | 12 | -20 | 3680 | 2 | 0 | 1824 | -4 | -4 |
| 88 | 12 | -16 | 3136 | 2 | 4 | 2928 | -4 | 0 |
| 96 | 10 | -16 | 512 | 2 | 12 | 1568 | -4 | 4 |
| 256 | 10 | -12 | 1392 | 2 | 8 | 768 | -4 | 8 |
| 208 | 10 | -8 | 96 | 2 | 16 | 64 | -4 | 12 |
| 24 | 8 | -16 | 16 | 2 | 24 | 56 | -4 | 16 |
| 256 | 8 | -12 | 8 | 0 | -16 | 192 | -6 | -12 |
| 736 | 8 | -8 | 768 | 0 | -8 | 688 | -6 | -8 |
| 228 | 8 | 0 | 1600 | 0 | -4 | 1664 | -6 | -4 |
| 576 | 8 | -4 | 4356 | 0 | 0 | 1248 | -6 | 0 |
| 192 | 6 | -12 | 2112 | 0 | 8 | 448 | -6 | 4 |
| 688 | 6 | -8 | 3264 | 0 | 4 | 128 | -6 | 8 |
| 1664 | 6 | -4 | 576 | 0 | 12 | 24 | -8 | -16 |
| 1248 | 6 | 0 | 120 | 0 | 16 | 256 | -8 | -12 |
| 448 | 6 | 4 | 64 | 0 | 20 | 736 | -8 | -8 |
| 128 | 6 | 8 | 2 | 0 | 32 | 576 | -8 | -4 |
| 96 | 4 | -12 | 64 | -2 | -12 | 228 | -8 | 0 |
| 704 | 4 | -8 | 624 | -2 | -8 | 96 | -10 | -16 |
| 1824 | 4 | -4 | 1920 | -2 | -4 | 256 | -10 | -12 |
| 2928 | 4 | 0 | 3680 | -2 | 0 | 208 | -10 | -8 |
| 1568 | 4 | 4 | 3136 | -2 | 4 | 32 | -12 | -20 |
| 768 | 4 | 8 | 1392 | -2 | 8 | 88 | -12 | -16 |
| 56 | 4 | 16 | 512 | -2 | 12 | 16 | -14 | -24 |
| 64 | 4 | 12 | 96 | -2 | 16 | 1 | -16 | -32 |
|   |   |   | 16 | -2 | 24 |   |   |   |

The results of partition function parameters calculations are in Tab. 1. The total accuracy of the computation was controlled by using of the total known number configurations, of the binomial coefficients (for 4x4 system there are 65536 configurations), also by other well-known parameters.

Other interesting outcome for given algorithm is in measuring of speedups which ones had super linear values for small number of processes. The running time of application was estimated with using of the library *time.h* and the MPI inline functions *clock(), MPI_Wtime(), ftime(),* system command "*time*" and duration writing of *datafile,* see Tab. 2.

TABLE 2. The measuring of process durations and speeding up by means of different methods for 5x5 lattice system. Time in sec.

|   |   | clock(); | MPI_Wtime(); | ftime(); | System Command "time" | open-closing datafile |
|---|---|---|---|---|---|---|
| The running time of the process # | 0 | 37.910 | 37.9297 | 38.662 | 38.613 | 37.478 |
|   | 1 | 37.460 | 37.4688 | 37.468 |   |   |
|   | 2 | 37.710 | 37.7188 | 37.66 |   |   |
|   | 3 | 37.530 | 37.918 | 38.612 |   |   |
| Single process |   | 156.280 | 156.312 | 156.315 | 156.536 | 156.286 |
| Speedup 4 nodes |   | 4.122 | 4.121 | 4.043 | 4.054 | 4.170 |
| Speedup 8 nodes |   | 8.068 | 8.064 | 8.062 | 7.714 | 8.08 |
| Speedup 16 nodes |   | 8.247 | 8.279 | 8.279 | 7.529 | 8.116 |

The cause of the strange and intriguing speedup values which ones are over maximal values (up to 4%) and in the contradiction with values predicted by the Amdahl's law, is not clear. The possible reason of such result could be quantities of the massive parallel logical-arithmetical operations over bit-vectors which ones could decrease because of cash hits are increases. The following decreases of the speedup are in connection with the growth of costs on message passing.

## 5. CONCLUSION

The high-performance parallel algorithm for rigorous calculation of partition function of the lattice systems of the finite number Ising spins was developed. The algorithm is scalable and it is possible to extend it for the running on huge calculation recourses. The facilities of calculation of 3D Ising lattices were elaborated. The nested iterations can be executed with utilization of the approach dynamical generation of the processes.

Unfortunately even with linear increasing of speedup by existing methods it is impossible to calculate $M_i$ and $E_i$ for each configuration of 10x10 system i.e. 100 spins in the given approach. So far as $2^{100}$ is approximately $10^{33}$, therefore for computational cloud with computer speed in order ExaMips ($10^{18}$ Mips) only for the summation until decillion it is necessary $10^{15}$ seconds or ~31,7 millions year! The same time needed for the 3D Ising lattice with structure 4x5x5. Of course, the availability of generating function could simplify the solution, but today we have not rigorous analytical solution of the simplest (!) ferromagnetism model even for 2D Ising lattice. The future approach to optimization of the chosen scheme of calculation of partition function could be done by means of taking into account the hypersymmetry of the this task. The solution obtained by authors can be used for comparison with experimental magnetization values for the finite atoms systems.